
\baselineskip=18pt plus 2pt minus 1pt
\magnification=1100
\hsize=6.0truein
\vsize=8.4truein
\voffset=24pt
\hoffset=.1in

\centerline{\bf SURPRISES ON THE WAY FROM 1D TO 2D QUANTUM MAGNETS:}
\centerline{\bf THE NOVEL LADDER MATERIALS.}

\vskip 1cm
\centerline{Elbio DAGOTTO$^1$ and T. M. RICE$^2$}
\vskip .15in
\centerline{\it $^1$Dept. of Physics and National High Magnetic Field Lab,}
\centerline{\it Florida State University, Tallahassee, FL 32306, USA}
\vskip .15in
\centerline{\it $^2$ AT\&T Bell Laboratories, Murray Hill, NJ 07974, USA, and}
\centerline{\it  Theoretische Physik, Eidgen\"ossische Technische
Hochschule,}
\centerline{\it 8093 Z\"urich, Switzerland}

\vskip 1cm

\centerline{\bf Abstract}

   One way of making the transition between the quasi-long range
order in a chain of S=1/2 spins coupled antiferromagnetically and
the true long range order that occurs in a plane, is by assembling
chains to make ladders of increasing width. Surprisingly this
crossover between one and two dimensions is not at all smooth.
Ladders with an even number of legs have purely short range
magnetic order and a finite energy gap to all magnetic excitations.
Predictions of this novel groundstate have now been verified
experimentally. Holes doped into these ladders are predicted to
pair, and possibly superconduct.

\vskip 1cm

\centerline{\bf I. Introduction}

\vskip 0.4cm

The unexpected discovery of high temperature superconductivity[1] in
lightly doped antiferromagnets has sparked renewed interest in low
dimensional quantum magnets. The parent cuprate insulators are now
considered the best examples of planar spin-1/2 antiferromagnets
with isotropic and predominantly nearest neighbor coupling. They show
simple long range antiferromagnetic (AF) order at low temperatures in agreement
with theory
which predicts an ordered ground state for the S=1/2 AF Heisenberg
model on a two dimensional (2D) square lattice.[2]
The one dimensional (1D) AF Heisenberg chain is also well understood.
A famous exact solution found by Bethe many years ago[3] showed that quantum
fluctuations prevent true long range AF order giving instead a slow
decay of the spin correlations essentially inversely with separation between
the spins.
Therefore it came as a great surprise when numerical calculations
found that the crossover from chains to square lattices, obtained
by assembling chains one next to the other to form ``ladders'' of
increasing width, was far from smooth. Although there is no
apparent source of frustration, quantum effects lead
to a dramatic dependence on the width of the ladder (i.e. the number of coupled
chains).

Ladders made from an $even$ number of legs have spin liquid ground
states so called because of their purely short range spin correlation.
An exponential decay of the spin-spin correlation is produced by a
finite $spin-gap$, namely a finite energy gap to the lowest S=1 excitation
in the infinite ladder. These even ladders may therefore be regarded as
realizations
of the unique coherent singlet ground state proposed some years ago
by Anderson in the context of the two dimensional S=1/2 AF Heisenberg
systems (the so-called Resonance Valence Bond (RVB) state).[4]

Ladders with $odd$ number of legs behave quite differently and display
properties similar to single chains at low energies i.e. gapless spin
excitations and a power-law falloff of the spin-spin correlations, apart from
logarithmic corrections. This
dramatic difference between even and odd ladders predicted by theory has
now been confirmed experimentally in a variety of systems.

2-leg S=1/2 ladders are found in
vanadyl pyrophosphate ${\rm (VO)_2 P_2 O_7}$ and in some cuprates e.g.
${\rm SrCu_2 O_3}$ (Fig.1) (here we use the convention that an ``m-leg ladder''
denotes m coupled  spin-1/2 chains).
Measurements of the spin
susceptibility show that it vanishes exponentially at low
temperature, a clear sign of a spin-gap. Neutron scattering and $\mu SR$
measurements are consistent with
short range spin order in the 2-leg ladders although, as we
stressed before, they are unfrustrated spin systems which classically
should order without a spin-gap. Further NMR measurements have
confirmed the large spin-gap in the excitation spectrum.

3-leg ladders (e.g. ${\rm Sr_2 Cu_3 O_5}$) by contrast show longer range
spin correlations and even true long-range order at low temperature due
to
weak interladder forces. There is excellent agreement between theory and
experiment confirming that there is a dramatic difference between even
and odd S=1/2 Heisenberg AF ladders.

Doped chains have long fascinated theorists because they form unusual
quantum liquids, so-called Luttinger liquids with many unique
properties.[5]
Although doping experiments in ladder compounds are just starting,
extensive theoretical studies have been made of doped
ladders. Again a clear difference between even and odd ladders is
predicted.
Even ladders are  specially interesting  because hole $pairing$ in a
relative ``d-wave'' state is found using a variety of
techniques which places  them in a different universality class of
one-dimensional systems than the Luttinger liquids found  in single chains
and odd ladders.

The next section reviews the theory of the S=1/2 AF Heisenberg
model on ladders. In Section III cuprates and other compounds
that are realizations of S=1/2 AF Heisenberg ladders are discussed
together with recent magnetic measurements.
Hole doping of ladders is the topic of Section IV with emphasis on
theoretical studies. Finally some
concluding  remarks are made in Section V.

\vskip 1cm

\centerline{\bf II. S=1/2 Heisenberg model on ladders: theoretical aspects.}

\vskip 0.4cm

  The properties of S=1/2 Heisenberg AF
models defined on 1D chains
or on 2D square lattices are well-known.
The model is defined by the Hamiltonian
$$
{\rm H = J \sum_{\bf \langle i,j \rangle } {\bf S}_{\bf i}.{\bf S}_{\bf
j} },
\eqno(1)
$$
\noindent where ${\bf i}$ is a vector labelling lattice sites
where spin-1/2 operators
${\rm {\bf S_{\bf i}} }$ are located.
${\rm {\bf \langle i,j \rangle } }$ denote nearest neighbors (n.n.)
sites. ${\rm J (>0)}$ is the antiferromagnetic exchange coupling that
provides the energy scale in the problem. This scale is material
dependent and it ranges from a few meVs to about 0.1eV in the case of
the high temperature superconductors. As explained above,
on 2D square lattices, the Heisenberg model has a
ground state with long range antiferromagnetic order, while in 1D chains
the spin-spin correlation decays slowly to zero as a power-law.
Both systems are spin $gapless$ i.e. there is no cost
in energy to create an excitation with S=1.

The new field of ladder systems started when Dagotto, Riera and
Scalapino[6] (see also Hirsch[7] and Dagotto and Moreo[8]) found
evidence
that 2-leg ladders have a finite $spin-gap$ i.e. a finite
energy is needed to create a S=1 excitation.
They started with
the simple limit obtained by generalizing Eq.(1) so that the
exchange coupling
along the rungs of a 2-leg ladder (denoted by ${\rm J'}$)
is much larger than the coupling J along the chains, ${\rm J' \gg J}$.
This idealization has the
advantage that
rungs interact only weakly with each other, and the dominant
configuration in the ground state is the product state with the spins on
each rung forming a spin $singlet$.
The energy in this limit
is approximately ${\rm E_{gs} = -
{{3}\over{4}} J' N}$, where ${\rm N}$ is the number of rungs and
${\rm -{{3}\over{4}} J'}$
the energy of each rung singlet state ${\rm | \psi \rangle_S }$ ${\rm [
= ( | \uparrow
\downarrow \rangle - | \downarrow \uparrow \rangle )/\sqrt{2} ] }$.
The ground state has total S=0,
since each rung is in a spin singlet. To produce
a S=1 excitation a rung singlet must be promoted to a S=1
triplet ${\rm | \psi \rangle_T = \{ | \uparrow \uparrow \rangle , \  ( |
\uparrow
\downarrow \rangle + | \downarrow \uparrow \rangle )/\sqrt{2} , \ |
\downarrow \downarrow \rangle \} }$.
An isolated rung-triplet has an energy ${\rm J'}$ above the rung
singlet. The coupling along the chains creates a band of S=1 magnons
with a dispersion law, ${\rm \omega(k) =  J' + J cos(k)} $ in the limit
${\rm J' \gg J}$. The spin-gap is the minimum excitation energy
${\rm \Delta_{spin} }= \omega(\pi) (\sim {\rm J' - J})$ which remains large in
this limit.[9]
Concurrently, the spins are mostly uncorrelated between rungs since
the spin correlations decay exponentially with distance along the chains
leading to the $spin$ $liquid$ nature of this state. Note, however, that the
spins are not disordered but in a unique isolated quantum-coherent groundstate.

In the other extreme, ${\rm J'/J=0}$, the two chains decouple but isolated
spin-1/2
Heisenberg $chains$ do not have a spin-gap and excitations with S=1
and wavevector
${\rm k=\pi}$ are degenerate with the ground state in the bulk
limit. To reconcile the different behavior in the limits ${\rm
J'/J \gg 1}$ and ${\rm J'/J=0}$, it was conjectured[6] that
the spin-gap should smoothly decrease
as ${\rm J'/J}$  is reduced, reaching ${\rm \Delta_{spin} = 0}$ at some
critical value of the coupling. Later,
Barnes, Dagotto, Riera and Swanson[9] observed that the power law decay
of the spin correlation in an isolated chain implies that a chain is in a
critical state and thus small perturbations can qualitatively alter
its properties. They predicted
that the spin-gap would vanish $only$ at ${\rm J'/J=0}$, so that
${\rm \Delta_{spin} > 0}$ at all ${\rm J'/J > 0}$
including the values of experimental interest, ${\rm J'/J \sim 1}$.
The ladder spin system would always be in a spin liquid state in
contrast to the more familiar cases of the 1D and 2D
Heisenberg models which are $gapless$.

Physical realizations of ladders like
${\rm Sr Cu_2 O_3}$ or ${\rm (VO)_2 P_2
O_7}$ correspond to ${\rm J' \approx J}$.
However, at ${\rm J'=J}$ there is no small parameter to guide a
perturbative calculation nor is an exact solution known.
$Numerical$ techniques can handle the region
${\rm J' \approx J}$ and
Exact Diagonalization of small clusters and Quantum Monte
Carlo techniques were used in Ref.[6,9] to study ${\rm \Delta_{spin}}$ as a
function of
${\rm J'/J}$.
The techniques used are not essential to
the discussion. We refer the reader elsewhere for
details,[10] and concentrate on the results.

In Fig.2a, ${\rm \Delta_{spin}}$
calculated  numerically by Barnes et
al.[9] shows that indeed ${\rm \Delta_{spin} > 0 }$
for all ${\rm J'/J \neq 0}$.
At the realistic coupling ${\rm J'=J}$, the gap is ${\rm \Delta_{spin}
\approx 0.5 J}$.
More recently,
White, Noack and Scalapino,[11] using a novel renormalization group (RG)
technique suitable
for static properties of 1D systems
reported a value ${\rm \Delta_{spin} = 0.504J }$ at ${\rm J'=J}$,
in excellent agreement with Refs.[6,9].
Note,
there are AF spin correlations at short distances along the chains
and across the
rungs, but even at ${\rm J'=J}$, the latter are somewhat stronger[11]
showing that the rough picture of a ground state dominated by
rung-singlets[6] is robust. Finally the closely related one-band Hubbard
model at half-filling also shows a spin-gap for all interaction
strengths.[12,13]

By now it is clear that
the presence of a spin-gap in the 2-leg ladder has been well
established using a variety of techniques. A useful intuitive
approximation is to visualize
the groundstate as mostly
rung singlets supplemented by weak AF
correlations along
the chains.
Gopalan, Rice and Sigrist[14] suggested that a good
variational description of the ground state of the 2-leg ladder could be
obtained using the short-range Resonance Valence Bond (RVB) state
proposed by Anderson and Kivelson et al.[4,15] with mostly adjacent rung
singlets, but including resonance between 2 adjacent rung singlets
into 2 n.n. singlets along the chains.[12]

What happens if we increase the number of ``legs'' in the ladder? This
is not a purely academic question since
Rice, Gopalan, and Sigrist[16] have shown
that materials like
${\rm Sr_{n-1} Cu_{n+1} O_{2n} }$ contain ladder
structures with a number of legs that depends
on the value of n.
The large ${\rm J'/J}$ limit
allows us again to make predictions for  the behavior of
the m-leg ladder. Let us begin with the $even$-leg ladder.
At ${\rm J'/J \gg 1}$, the rungs
decouple and at each level one has ${\rm 2^m}$ states instead
of the four states of the 2-leg ladder, apparently complicating the problem.
However, the ground state of the m-spins rung is
also a S=0 singlet
separated by a finite gap from the first excited state. Thus, as in the
case of the 2-leg ladder,
the even-leg ladder at ${\rm J' \gg J}$ has a finite spin-gap
${\rm \propto J'}$. Taking the analogy with the 2-leg ladder further
it is again plausible to assume that a spin-gap  exists in the
even-leg ladder for any ${\rm J'/J \neq 0}$.
Early numerical calculations on 4-leg ladders are in
agreement with this picture.[8]
Poilblanc  et al.[17] evaluating exactly the $4 \times 6$ and $4 \times 8$
clusters with periodic boundary conditions, and extrapolating the results
to the bulk limit using an exponential form obtain
a spin-gap ${\rm \Delta_{spin} = 0.245J}$, about half the
size of the gap for the 2-leg ladder.  Hatano and Nishiyama[18] found ${\rm
\Delta_{spin} = 0.27J}$ using a similar analysis.
A reduction in the size of the gap  is
natural since as the width of the ladders
grows, the 2D square lattice limit is approached and ${\rm \Delta_{spin}
\rightarrow 0}$.
White et al.[11] using a RG technique on larger ${\rm 4 \times N}$ clusters but
with open boundary conditions, which amplify the finite size effects,
reported ${\rm \Delta_{spin} = 0.190J}$ extrapolated to ${\rm N
\rightarrow \infty}$, with a spin correlation
length $\xi_{AF} \sim 5-6$.
Finally, a mean
field approach[14] predicted ${\rm \Delta_{spin} = 0.12J}$.
The presence of a finite spin-gap in
the 4-leg ladder seems by now well established theoretically, but some
discrepancies on its value remain to be clarified.

  Rice et al.[16] and Gopalan et al.[14] quoting arguments by Hirsch and
Tsunetsugu made the interesting observation that
ladders with an odd number of legs should behave quite differently from
even-leg
ladders and display
properties similar to single chains at low energies i.e. $gapless$ spin
excitations and a power-law falloff of the spin-spin correlations.
The simplest way to visualize this difference
is again by analyzing the large ${\rm J'/J}$
limit, as remarked by Reigrotzki et al.[19]
Let us consider for example the 3-leg ladders. At large
${\rm J'/J}$, each rung can be diagonalized exactly leading to a $doublet$
ground
state, and a doublet and quadruplet excited states. The rung
doublet of lowest energy will be the dominant configuration in the ground state
at small
temperature which thus consists now of S=1/2 states (doublets) in each rung.
The inter-rung coupling J provides with an effective interaction
between these
S=1/2 rung states which by rotational invariance must be of the Heisenberg form
with an effective coupling ${\rm J_{eff}}$ as energy scale.
Thus, the ground state properties of
the 3-leg ladder at large ${\rm J'/J}$ should be those of the spin-1/2
Heisenberg
chain, with a coupling ${\rm J_{eff}}$ instead of ${\rm J}$, and thus
with a
vanishing spin-gap. The argument can be trivially generalized to all
odd-leg ladders. Since for the odd-leg case
both at ${\rm J'/J \gg 1 }$ and ${\rm J'/J=0}$ there is no spin-gap, it
is reasonable that at intermediate values of ${\rm J'/J}$ the
gap always vanishes in contrast to even-leg ladders.
A recent numerical  RG calculation[11] verified these intuitive
ideas.

Khveshchenko[20] explained the qualitative difference between even and
odd ladders based on an argument used by Haldane for the 2D square
lattice.[21]
For odd ladders a topological term governing the dynamics at
long-wavelengths appears in the effective action, while for even ladders it
exactly cancels. This topological term is similar to the one  that causes the
well-known
difference between the finite spin-gap of  integer Heisenberg spin chains and
the
absence of a spin-gap for half-integer spin chains.
The direct analogy  with the Haldane state of the S=1 chain is
realized in ladders with a $ferromagnetic$ coupling, ${\rm J' < 0}$,
on the rungs. In this case in the ${\rm |J'/J| \gg 1}$ limit the rungs become
spin triplets rather than singlets, and the relation with the S=1
chain is obvious.[22] However, more work is needed to clarify the relationship
of the
Haldane state of the integer spin chains and the spin liquid of the AF
spin-1/2 even-leg ladders.[23]

The single magnon spectrum ${\rm \omega(k)}$ of the 2-leg ladder evolves from
a simple cosine dispersion at ${\rm J' \gg J}$, dominated
by S=1 rung states,[9] to a more linear dispersion
around the minimum at ${\rm \omega(\pi) \approx 0.5J}$ at the isotropic
coupling value, ${\rm J' = J}$ (Fig.2b).[9,14] This change can be traced to a
spreading
of the two parallel spins in the triplet over more than one rung as
${\rm J'/J}$ is reduced, which in turn modifies the dispersion relation
through longer range transfer processes. As shown by  Barnes and
Riera [24], the magnons near ${\rm k=\pi}$ remain as well
defined modes separated from the 2-magnon continuum which starts at
energy ${\rm \approx J}$ near ${\rm k=0}$.[25,26]  The magnon dispersion should
in principle be directly measurable through inelastic neutron scattering
experiments on single crystals but as discussed below only powder
spectra are available at present.

Recently, thermodynamic properties of S=1/2 ladders have also been studied
by several groups. Troyer, Tsunetsugu and W\"urtz[27] using a quantum
transfer-matrix method on 2-leg ladders obtained reliable results down
to temperature ${\rm T \approx 0.2J}$.
The correlation length $\xi_{AF}$ of the short range AF order values
$\xi_{AF} \approx 3-4$ (in units of the lattice spacing),
in agreement with calculations at zero temperature[11]
that reported $\xi_{AF} =3.19$.
The magnetic susceptibility $\chi(T)$ [24,27,28]
crosses over from a Curie-Weiss form ${\rm \chi(T) = C/(T + \theta) }$ at
high temperature, to an exponential falloff ${\rm \chi(T) \sim }$
${\rm exp({- {{\Delta}_{spin}/{T}} })/\sqrt{T} }$ as ${\rm T \rightarrow 0}$
reflecting the finite spin-gap.[29]
Recently, Frischmuth, Troyer and W\"urtz[30], using an improved algorithm,
extended their results to lower temperatures and to ladders up to 6-legs
in width. Their results for $\chi(T)$ are shown in Fig.3. The difference
between odd and even ladders is very clear at low temperatures as is the
smaller spin-gap in the 4- and 6-leg ladders.

\vskip 1cm

\centerline{\bf III. Experimental results on ladder compounds.}

\vskip 0.4cm

At present two types of ladder compounds are known. The first to be
identified was vanadyl pyrophosphate ${\rm (VO)_2 P_2 O_7}$ whose
structure was shown in Fig.1. The V-ions are in an oxidation state
${\rm V^{4+} }$ i.e. ${\rm 3d^1}$ with the single electron
occupying a non bonding ${\rm t_{2g} }$-orbital. The superexchange
interaction occurs through the ${\it dp\pi}$-overlap of V 3d- and
O 2p-orbitals. The magnetic susceptibility $\chi(T)$ shown in Fig.4a was
measured
by Johnston et al.[31] who found an activated behavior at
temperatures ${\rm T < 100K}$ crossing over to Curie-Weiss form
at higher temperature. Barnes and Riera [24] by fitting $\chi(T)$,
found almost equal values ${\rm J=7.76}$ meV and ${\rm J'=7.8}$ meV
for the exchange along the chains and rungs, respectively. Recently
Eccleston et al.[32] used a powder time-of-flight neutron
scattering technique to obtain the inelastic spectrum. The powder
average of the dynamic magnetic structure factor (Fig.5) shows
clear evidence of a spin-gap, ${\rm \Delta_{spin} \approx 3.7 \pm 0.2}$ meV
at a wavevector $\pi$, a value which agrees
well with the theoretical prediction of ${\rm \Delta_{spin} = 0.5 J =
3.9 meV}$.[6,9]
The data do not allow a unique determination of the magnon dispersion
relation but are consistent with the form illustrated in Fig.2b.

The second type of ladder compounds are cuprates but with
modified copper-oxygen planes and other structures. The key point here is the
configuration
of the ${\rm Cu O_4 }$ squares. The high-Tc
cuprate families all are based on ${\rm Cu O_2 }$-planes with
all corner sharing ${\rm Cu O_4}$-squares. This leads to $180^o$
${\rm Cu-O-Cu}$ bonds. Since the ${\rm Cu^{2+}}$-ion has a
$3d^9$ configuration with the single hole occupying an antibonding
$e_g$-orbital, there is an exceptionally strong superexchange
interaction (${\rm J \approx 0.13 eV}$) through $dp\sigma$ overlap
with the O2p-orbital common to both ${\rm CuO_4}$-squares.
In the ideal ${\rm CuO_2}$-plane, the ${\rm O}$-ions form a square
lattice and the ${\rm Cu}$-ions occupy the centers of exactly one
half of the ${\rm O_4}$-squares, also forming a square lattice. If
a line defect is introduced in the ${\rm Cu}$-occupation so that left
and right different sets of ${\rm O_4}$-squares are occupied, then along
this line the coordination of the ${\rm CuO_4}$-squares is edge-sharing
(Fig.1b).
But the superexchange path for two ${\rm CuO_4}$ squares sharing an edge
is very different and involves primarily an intermediate state with
2 holes on orthogonal orbitals on the same O-ion. Hund's Rule then
favors parallel spin alignment and as a result the Kanamori-Goodenough
rules give a weak ferromagnetic (F) coupling between ${\rm Cu^{2+}}$-ions
which are edge-sharing.

Hiroi, Azuma, Takano, and Bando,[33] were the first to synthesize the family of
layer compounds ${\rm Sr_{n-1} Cu_{n+1} O_{2n} }$ which have arrays
of parallel line defects. The copper oxide planes in the first two
members were shown in Fig.1b. Rice et al.[16] pointed out that nearly ideal
ladder compounds should result, since the pattern of strong AF $180^o$
${\rm Cu-O-Cu}$ bonds make a ladder, and the interladder coupling
is very weak both because of weak F $90^o$ ${\rm Cu-O-Cu}$ bonds and
the resulting frustration. The first member (n=3 or ${\rm Sr Cu_2 O_3}$)
has 2-leg ladders, the second (n=5 or ${\rm Sr_2 Cu_3 O_5}$) has 3-leg
ladders and so on.

Recently Azuma et al.[34] reported magnetic susceptibility measurements
for the 2- and 3-leg ladder compounds (see Fig.4b,c). The difference
between the two compounds is striking. The spin-gap is clearly
visible in the precipitous drop in $\chi(T)$ for ${\rm T < 300K}$ in the
2-leg compound, and by fitting to the low temperature form ${\rm \chi(T)
\sim T^{-1/2} exp(-\Delta_{spin} /T) }$, they obtained a value
${\rm \Delta_{spin} = 420K}$. This compound should have exchange constants
close to
the isotropic limit ${\rm J = J' \approx 1300K}$ so that theory
predicts a larger value for an isolated ladder ${\rm \Delta_{spin}^{theory}
\sim
650K}$. However,
in ${\rm Sr Cu_2 O_3}$ there is substantial exchange coupling along the
c-axis, ${\rm J_c}$. This should lower ${\rm \Delta_{spin}}$ and may account
for most of the discrepancy. But, Azuma et al.[34] (see also Ishida
et al.[35]) also reported NMR investigations. In particular, they observed
activated behavior in the relaxation rate ($1/T_1$) at ${\rm T < 300K}$
as expected but the activation
energy (680K) was substantially larger than the value deduced from
$\chi(T)$. At present the origin of the discrepancy is unclear.

Azuma et al.[34] measured $\chi(T)$ also for the 3-leg ladder
compound, ${\rm Sr_2 Cu_3 O_5}$, and found $\chi(T) \rightarrow const.$
as $T \rightarrow 0$ as expected for the 1D AF Heisenberg chain. Further
$\mu SR$ measurements by Kojima et al.[36] found evidence of a long range
ordered state with ${\rm T_N = 52 K}$. This ordering we attribute to the
interlayer coupling ${\rm J_c}$ along the c-axis. Note that no sign of
long-range
ordering was observed in the 2-leg compound. These results confirm
explicitly the drastic difference between ladders with even and odd
number of legs.

Ladder structures occur also in other cuprates. For example, Batlogg
et al.[37] reported the magnetic susceptibility for the family of
compounds ${\rm La_{4+4n} Cu_{8+2n} O_{14+8n} }$
which, as special cases, contain 4- and 5-leg ladder
elements. These are complex structures which
contain other Cu-sites
only weakly coupled to each other. These latter spins
dominate $\chi(T)$ below room temperature.
However, by examing the difference in $\chi(T)$ between the two
compounds, Batlogg et al.[37] could identify a substantial spin-gap
in a  4-leg compound (${\rm \Delta_{spin} \approx 300K}$).
Note that in these compounds only weak inter-ladder coupling  is
expected and a value
of ${\rm \Delta_{spin} \approx J/4 ( \approx 325 K)}$ is predicted
theoretically, which agrees quite well with the experiment. Very
recently, Hiroi and Takano [38] have found a new ladder compound ${\rm
La Cu O_{2.5}}$ with 2-leg ladders which are weakly connected in a three
dimensional structure.

\vskip 1cm

{\bf IV. Hole doping of spin-1/2 ladders}

\vskip 0.4cm

Generally it is difficult to dope transition metal oxides and produce a
highly conducting state but the cuprates are exceptional in this regard.
Early reports of doped cuprate ladder materials are starting to appear.[38,39]
Apart from the possibility of realizing doped ladders, their behavior is
of great interest to theorists because they are examples of unusual
Fermi liquids that can be carefully analyzed. Hole doping of a cuprate
introduces effective ${\rm Cu^{3+}}$-sites. This oxidation state also favors
square planar O-coordination similar to ${\rm Cu^{2+}}$-ions and in this
coordination a low spin S=0 ${\rm Cu^{3+}}$-ion is formed which
corresponds to a bound state of a S=1/2 ${\rm Cu^{2+}}$-ion
and a hole residing
mainly on the four surrounding O 2p-orbitals (Zhang-Rice singlet).[40]
Transfer of electrons between n.n. sites allows a S=1/2 ${\rm Cu^{2+}}$ and
S=0 ${\rm Cu^{3+}}$-ion to exchange positions. The canonical model
describing the motion of the effective S=0 ${\rm Cu^{3+}}$-ions in a background
of Heisenberg coupled S=1/2 ${\rm Cu^{2+}}$-ions is known as the ${\rm
t-J}$ model.[41]

The properties of a hole doped single chain have been much studied. It
is an example of a Luttinger liquid - so called to distinguish it from
the Landau Fermi liquid state that is ubiquitous for interacting
fermions at low temperatures in higher dimensions. The simple
alternating AF spin pattern of the parent insulator changes its period
to an incommensurate value which depends on the doping. The exponent of
the power-law decay increases but magnetic correlations still are the
dominant ones. The most striking feature of Luttinger liquids is
spin-charge separation whereby the charge and spin parts of an added
hole move at different velocities and become spatially separated
from each other.[5] All these properties are fascinating but do not give a
sign of impending superconductivity.

The 2-leg ladder starts from a very different parent state characterized
by a spin-gap and exponentially decaying spin correlations. A key
question is how these features evolve with doping. Mean-field studies by
Sigrist, Rice and Zhang[42] found an increase in
the gap upon doping but numerical studies of finite length ladders
by Dagotto et al.[6], Poilblanc et al.[17,43]
and Noack et al.[12]
found a decrease. Detailed studies by Tsunetsugu et al. [44] showed
that it was necessary to distinguish two different types of magnetic
excitations. Again the limit ${\rm J' \gg J}$ is useful to gain
intuition. As remarked by Dagotto et al.[6], in this limit
holes $pair$ on the same rung in a S=0 and zero momentum state
to reduce the cost in
magnetic interactions. One type of magnetic excitation is to promote a
singlet pair of spins spatially separated from the hole pairs to form a S=1
triplet and this excitation evolves smoothly from the magnon we
discussed earlier in the undoped case. However, a new type of spin
excitation is now possible.[44] This involves separating the hole pair into
a state with the holes on two spatially separated rungs, each of which
now contains an unpaired spin. This new excitation still requires a
finite energy so the spin-gap and the exponential decay of the spin-spin
correlations remain, but its appearance at a new and lower energy than
the magnon mode leads to a discontinuity in the spin-gap upon doping. Note,
since these excitations require holes their number vanishes as the undoped
insulator is approached.

The early calculations of Dagotto et al.[6]  supported a continuous
evolution of the doped system from the anisotropic limit ${\rm J' \gg
J}$ where strong pairing correlations signaling superconductivity were
observed, down
to the isotropic case ${\rm J' = J}$. This is similar to the smooth connection
observed in the undoped Heisenberg models. The mean field calculations of
Sigrist et al.[42] were in agreement and further lead to the observation
that holes were paired in a state of approximate d-wave symmetry
although the lack of rotational invariance of the lattice here prevents an
exact
symmetry classification. Calculations by Tsunetsugu et al.[45]
confirmed that this `d-wave' paired state for holes persisted down to
the limit ${\rm J' = J}$ and realistic (for cuprates)
values of ${\rm J/t \sim 0.3}$.[46,47] The
size of the hole pair is now larger than a single rung but they are
spread only over a few lattice spacings. The excitation spectrum of the doped
2-leg ladder contrasts with the Luttinger liquid in that the ladder low energy
sector contains only the collective sound mode of the bosonic liquid of
hole pairs and a finite energy is needed to make a triplet excitation.
These features are similar to the case of $attractive$ fermions in a
single chain as analyzed by Luther and Emery [48] rather than the repulsive
case which is the Luttinger liquid discussed earlier. Another
interesting
aspect of lightly doped 2-leg ladders is the way in which they combine
features of lightly doped insulators with those of metals with large
Fermi surfaces. The former behavior dominates in the energy dependence
of the spectral function to add electrons but metallic behavior appears
in the momentum dependence of added quasiparticles.[26,45,49]

Binding hole pairs gives them a bosonic character which in turn is a
necessary step on the way to superconductivity. However, this alone does
not suffice since a groundstate with a crystalline order of hole pairs is also
possible.[6] Actually in a quasi-one-dimensional system like a ladder, true
long range order will be prevented by quantum fluctuations but a
power-law falloff will persist. In the doped ladder this occurs both
in the channels corresponding to crystalline ordering of hole pairs and
that with superfluid or Bose condensation of hole pairs. The balance
between the two and the question of which dominates by means of a
smaller exponent depends on the parameters of the model and more
generally on residual interactions between hole pairs. This is hard to
predict accurately.[45,50] The first set of experiments by Hiroi and Takano
[38]
on ${\rm La_{1-x} Sr_x Cu O_{2.5} }$, a doped 2-leg
ladder system, show substantial decreases in the resistivity upon doping
and evidence of metallic behavior in resistivity vs temperature at the
highest value of ${\rm x=0.2}$ (see Fig.6). There are signs that the spin-gap
persists upon doping at least initially but there are no signs of
superconductivity. More experiments will be needed to determine if hole
pairing exists and if the disorder is suppressing the superconductivity.
Nonetheless conceptually the relation of the paired hole
state of the doped 2-leg ladder to the superconducting state of the planar
cuprates is much closer than the relation to  the single chain or Luttinger
liquid state.

\vskip 1cm

\centerline{\bf V. Conclusions}

\vskip 0.4cm

The study of low dimensional quantum antiferromagnets has emerged as a
central problem in condensed matter physics due to the discovery of
high-Tc superconductivity in lightly doped cuprates with planar
structures. Quantum effects are largest in a S=1/2 system and with
isotropic Heisenberg coupling. A square lattice still has an
ordered groundstate although with a substantial reduction of the
sublattice magnetization due to quantum effects. In the one dimensional
analog, i.e. a Heisenberg S=1/2 chain, the quantum effects overwhelm the
long range order but the groundstate has quasi long-range order with a
decay in the spin-spin correlation function as an inverse power in the
separation,
apart from logarithmic corrections.

One might expect that a two leg ladder should be intermediate between a chain
and a plane
thus the discovery that quantum effects are much stronger in such a ladder
and lead to purely short range order with an  exponential decay in
spin-spin correlations came as a great surprise. This result first found
in numerical simulations has now been verified by a variety of
techniques and more importantly has experimental confirmation in
${\rm (VO)_2 P_2 O_7}$, ${\rm SrCu_2 O_3}$, and ${\rm La Cu O_{2.5} }$.This
difference between a single chain and two-leg ladder extends to all odd-
and even-leg ladders,  and the difference can be
traced to the absence in even-leg ladders of the special topological
term that appears in the low energy action of the single chain. This term
is also absent in integer spin chains and these also
display exponentially decaying spin-spin correlations and a spin-gap.

The various families of high-Tc superconductors all have a unique
structural
element, namely ${\rm Cu O_2}$-planes composed of a square lattice of Cu-ions
separated by O-ions. The local coordination is characterized by ${\rm Cu
O_4}$-squares which in turn are all corner sharing in the ${\rm Cu
O_2}$-planes. The ladder cuprates again have the same local ${\rm Cu
O_4}$-coordination but the pattern of the ${\rm Cu
O_4}$-squares is changed which in turn changes the pattern of magnetic exchange
interactions. For example, in ${\rm Sr_{n-1} Cu_{n+1} O_{2n}}$ line
defects break the plane up into weakly coupled ladders. The many ways of
assembling ${\rm Cu O_4}$-squares
illustrates the richness of cuprate chemistry which is only now
beginning to be explored and various possibilities for novel quantum
groundstates remain to be studied.

The cuprates have another unique feature among transition metal oxides,
namely the possibility of hole doping without localization to realize
conducting materials.
The doped chain has been the paradigm of a non-Landau Fermi liquid and
much attention has focussed on the unique properties of this quantum
liquid, called a Luttinger liquid by Haldane, such as the complete
separation of charge and spin sectors into two excitation branches at
low energy. The hole doped 2-leg ladder is also essentially one
dimensional but now the properties are radically different. As we
discussed above, the quantum liquids in lightly doped ladders retain the
spin-gap, show hole-hole pairing in approximate ${\rm d_{x^2 -
y^2}}$-symmetry, and although lightly doped insulators they show features of a
large Fermi surface which is metal-like. Doped ladders are a
fascinating mixture
of a dilute Fermi gas with strong attractions and a concentrated Fermi
system with a large Fermi surface.

Returning to the high-Tc cuprates we see a paradox. The parent
insulating antiferromagnets show long range order, which represents a
smooth evolution or crossover from the properties of single chains but
not from 2-leg ladders. Lightly doped cuprates by contrast show a spin
gap and ${\rm d_{x^2 - y^2} }$-superconductivity, properties we can
imagine evolving smoothly from the 2-leg ladders. While much remains to
be done to understand how these features fit together, it is clear that
the study of ladders has given us not only surprises but valuable new
insights into low dimensional quantum systems and a new impetus to broaden
our horizons  and explore the rich solid state chemistry of cuprates and
related materials.

\vskip 0.5cm

\centerline{\bf Acnowledgments}

We thank M. Azuma, T. Barnes, H. Hiroi, A. Moreo,
J. Riera, D. Scalapino, M. Takano,
M. Troyer, and H. Tsunetsugu
for their help in the preparation of this review.
E.D. specially thanks
the Office of Naval Research under
grant ONR N00014-93-0495 for its support. He
also thanks the NHMFL and MARTECH for additional support.

\vfill\eject

\centerline{FIGURE CAPTIONS}

Fig.1: (a) Ladder compound ${\rm (VO)_2 P_2 O_7}$. O- and
V-ions are indicated (from D. C. Johnston et al., Phys. Rev. B{\bf 35},
219 (1987)); (b) Schematic representation of the
2-leg compound ${\rm Sr Cu_2 O_3}$ and the 3-leg compound ${\rm Sr_2
Cu_3 O_5 }$ (from M. Azuma et al., Phys. Rev. Lett. {\bf 73}, 3463 (1994)).
The black dots are copper, while the intersections of the solid lines
are oxygen locations. The dashed lines are Cu-O bonds. The 2- and 3-leg
structures are highlighted. ${\rm J}$ is the coupling along the chains,
and ${\rm J'}$ along the rungs.

Fig.2: (a) Spin gap ${\rm \Delta_{spin}}$ vs. ${\rm J'/J}$ for
the 2-leg ladder. The results are extrapolations to the bulk limit
using numerical results obtained on finite ${\rm 2 \times N}$ clusters
(from T. Barnes et al., Phys. Rev. B{\bf 47}, 3196 (1993)); (b) Triplet
spin-wave excitation spectra for the isotropic point ${\rm J'=J}$,
with ${\rm J=7.79}$ meV and using the ${\rm (VO)_2 P_2 O_7}$ lattice spacing
(from
T. Barnes and J. Riera, Phys. Rev. B{\bf 50}, 6817 (1994)). ${\rm k_r}$
(${\rm k_c}$) denotes momentum along the rungs (chains).

Fig.3: Magnetic susceptibility $\chi(T)$ calculated with Monte Carlo
techniques on m-leg ladders and ${\rm J'=J}$ on clusters of ${\rm m
\times 100}$ sites. $\chi(T)$ for  even-leg ladders show at low
temperature the exponential
suppression caused by the spin-gap, while the odd-leg ladders extrapolate
to a finite number as ${\rm T \rightarrow 0}$ (from B. Frischmuth, M. Troyer
and D. W\"urtz, preprint).

Fig.4: (a) Experimental magnetic susceptibility $\chi(T)$ for ${\rm (VO)_2
P_2 O_7}$ (from D. C. Johnston et al., Phys. Rev. B{\bf 35},
219 (1987)); same but for (b) ${\rm Sr Cu_2 O_3}$ and (c) ${\rm Sr_2
Cu_3 O_5 }$ (from M. Azuma et al., Phys. Rev. Lett. {\bf 73}, 3463 (1994)).

Fig.5: Neutron scattering data for ${\rm (VO)_2 P_2 O_7}$ showing the finite
spin-gap $E_g$ (from
R. S. Eccleston et al., Phys. Rev. Lett. {\bf 73}, 2626 (1994)).
We refer the reader to this reference for further details.

Fig.6: (a) Resistivity vs temperature parametric with the Sr-concentration $x$
for
${\rm La_{1-x} Sr_x Cu O_{2.5} }$; (b) Magnetic susceptibility vs temperature
for
the same compound shown in (a) (from Z. Hiroi and M. Takano,
Nature Vol.{\bf 377}(7), 41 (1995)).

\vfill\eject

\centerline{REFERENCES}
\medskip

\item{1} J. Bednorz and K. M\"uller,
Z. Phys. {\bf B 64}, 188 (1986).


\item{2} E. Manousakis, Rev. Mod. Phys. {\bf 63}, 1 (1991).

\item{3} H. Bethe, Z. Phys. {\bf 71}, 205 (1931).

\item{4}  P. W. Anderson, Science {\bf 235}, 1196 (1987).

\item{5} For a recent review see H. J. Schulz, {\it Proceedings of Les Houches
Summer School LXI: Mesoscopic Quantum Physics}, Eds. E. Akkermans,
G. Montambaux, J.L Pichard, and J. Zinn-Justin, Elsevier,
Amsterdam, to be published; and references therein.

\item{6} E. Dagotto, J. Riera, and D. Scalapino, Phys. Rev.
B{\bf 45}, 5744 (1992).

\item{7} R. Hirsch, Diplomarbeit, Universit\"at K\"oln (1988).

\item{8} Dagotto and Moreo, Phys. Rev. B{\bf 38}, 5087 (1988).

\item{9} T. Barnes, E. Dagotto, J. Riera and E. Swanson,
Phys. Rev. B{\bf 47}, 3196 (1993).

\item{10} E. Dagotto, Rev. Mod. Phys. {\bf 66}, 763 (1994).

\item{11} S. White, R. Noack, and D. Scalapino, Phys. Rev. Lett.
{\bf 73}, 886 (1994).

\item{12} R. Noack, S. White, and D. Scalapino, Phys. Rev. Lett.
{\bf 73}, 882 (1994).

\item{13} M. Azzouz, L. Chen, and S. Moukouri, Phys. Rev. B{\bf 50},
6233 (1994).

\item{14} S. Gopalan, T. M. Rice and M. Sigrist, Phys. Rev.
B{\bf 49}, 8901 (1994).

\item{15} S. A. Kivelson, D. S. Rokhsar, and J. P. Sethna, Phys. Rev.
{\bf B 35}, 8865 (1987).

\item{16} T. M. Rice, S. Gopalan, and M. Sigrist, Europhys. Lett.
{\bf 23}, 445 (1994). See also H. J. Schulz, Phys. Rev. {\bf B 34}, 6372
(1986); I. Affleck, Phys. Rev. {\bf B 37}, 5186
(1988); D. S. Rokhsar and S. A. Kivelson, Phys. Rev. Lett. {\bf 61},
2376 (1988); N. E. Bonesteel, Phys. Rev. {\bf B 40}, 8954 (1989); and
references therein.

\item{17} D. Poilblanc, H. Tsunetsugu, and T. M. Rice,
Phys. Rev. B{\bf 50}, 6511 (1994).

\item{18} N. Hatano and Y. Nishiyama, to be published in J. Phys. A:
Math. Gen.

\item{19} M. Reigrotzki, H. Tsunetsugu, and T. M. Rice, J. Phys. C:
Cond. Matt. {\bf 6}, 9325 (1994).

\item{20} D. V. Khveshchenko, Phys. Rev. B{\bf 50}, 380
(1994) and D. G. Shelton, A. A. Nersesyan and A. M. Tsvelik, preprint.

\item{21} F. D. M. Haldane, Phys. Rev. Lett. {\bf 61}, 1029 (1988).

\item{22}  For work with ferromagnetic rungs see
K. Hida, J. Phys. Soc. Jpn. {\bf 60}, 1347 (1991),
H. Watanabe, Phys. Rev. {\bf B 50}, 13442 (1994), and
references therein.

\item{23}  S. P. Strong and A. J. Millis, Phys. Rev. Lett. {\bf 69}, 2419
(1992); Phys. Rev. {\bf B 50}, 9911 (1994);
Y. Xian, Manchester preprint;
Y. Nishiyama, N. Hatano and M. Suzuki, to appear in J. Phys.
Soc. Jpn. {\bf 64}, No.6, (1995); S. R. White, preprint;
H. Watanabe, preprint;
D. S\'en\'echal, preprint.

\item{24} T. Barnes and J. Riera, Phys. Rev. B{\bf 50}, 6817 (1994).

\item{25} The dynamical spin structure factor
$S({\bf Q},\omega)$, with ${\bf Q}=(\pi,\pi)$ has a sharp peak
at ${\rm \omega =
\Delta_{spin} \approx 0.5J}$ that carries most of the weight. See Refs.[9,14].

\item{26} M. Troyer, H. Tsunetsugu and T. M. Rice, ETH-preprint.

\item{27} M. Troyer, H. Tsunetsugu, and D. W\"urtz, Phys. Rev. {\bf B
50}, 13515 (1994).

\item{28} A. W. Sandvik, E. Dagotto, and D. J. Scalapino, preprint.

\item{29} In addition, the nuclear spin relaxation rate has been calculated by
Troyer
et al.[27] predicting ${\rm 1/T_1 \sim exp({- {{\Delta_{spin}}/{T}}}) (a +
lnT)}$  at low temperatures.
Sandvik et al.[28] have also calculated ${\rm 1/T_1}$ using numerical
techniques.

\item{30} B. Frischmuth, M. Troyer and D. W\"urtz, preprint.

\item{31} D. C. Johnston, J. W. Johnson, D. P. Goshorn and
A. P. Jacobson, Phys. Rev. B{\bf 35}, 219 (1987).

\item{32} R. S. Eccleston, T. Barnes, J. Brody and J. W.
Johnson, Phys. Rev. Lett.{\bf 73}, 2626 (1994).

\item{33} Z. Hiroi, M. Azuma, M. Takano and Y. Bando, J. Sol. State Chem.
{\bf 95}, 230 (1991).

\item{34} M. Azuma, Z. Hiroi, M. Takano, K. Ishida and Y. Kitaoka,
Phys. Rev. Lett. {\bf 73}, 3463 (1994).

\item{35} K. Ishida, Y. Kitaoka, K. Asayama, M. Azuma, Z. Hiroi and M. Takano,
J. Phys. Soc. Japan {\bf 63}, 3222 (1994); K. Ishida, Y.
Kitaoka, Y. Tokunaga, S. Matsumoto, K. Asayama, M. Azuma, Z. Hiroi and
M. Takano, preprint.

\item{36} K. Kojima, et al., Phys. Rev. Lett. {\bf 74}, 2812 (1995).

\item{37} B. Batlogg et al., Bull. Am. Phys. Soc. {\bf 40}, 327 (1995).

\item{38} Z. Hiroi and M. Takano, Nature Vol.{\bf 377}(7), 41 (1995).

\item{39} M. Azuma, M. Takano, T. Ishida, and K. Okuda, preprint,
recently presented
magnetic susceptibility measurements for Zn-doped 2- and 3-leg ladders.

\item{40} F. Zhang and T. M. Rice, Phys. Rev. {\bf B 37}, 3759 (1988).

\item{41} Note that in the case of
${\rm (VO)_2 P_2 O_7}$, it is not clear whether the ${\rm t-J}$ model would be
suitable for describing its properties upon doping. A detailed
analysis based on a many orbital Hubbard model for V- and O-ions is needed.

\item{42} M. Sigrist, T. M. Rice and F. C. Zhang, Phys. Rev.
B{\bf 49}, 12058 (1994).

\item{43} D. Poilblanc, D. J. Scalapino, and W. Hanke, Phys. Rev. B (in press).

\item{44} H. Tsunetsugu, M. Troyer and T. M. Rice, Phys. Rev.
B{\bf 49}, 16078 (1994).

\item{45} H. Tsunetsugu, M. Troyer, and T. M. Rice, Phys. Rev. B{\bf
51}, 16456 (1995). See
also J. A. Riera, Phys. Rev. B {\bf 49}, 3629 (1994).

\item{46} C. A. Hayward, D. Poilblanc, R. M. Noack, D. J. Scalapino and
W. Hanke, Phys. Rev. Lett. {\bf 75}, 926 (1995).

\item{47} Calculations of superconducting correlations for the
Hubbard model were discussed in Ref.[12] and also in
Y. Asai, Phys. Rev. B{\bf 50}, 6519 (1994), and preprint;
Yamaji and Y. Shimoi, Physica {\bf C 222}, 349 (1994); and
R. M. Noack, S. R. White
and D. J. Scalapino, Europhys. Lett. {\bf 30}, 163 (1995), and preprint.

\item{48} A. Luther and V. J. Emery, Phys. Rev. Lett. {\bf 33},
589 (1974).

\item{49} Note the  similarities between the
results for 2-leg ladders and those corresponding to models of high-Tc
superconductors where the infrared Drude weight scales as the hole
concentration,
while the Fermi surface is large and
electron-like.[see Ref.10]

\item{50} Several recent publications have addressed these issues:
N. Nagaosa (Univ. of Tokyo preprint) claimed that the low energy physics of
doped
ladders is described in terms of bipolarons.
L. Balents and M. P. A. Fisher (ITP preprint) studying the Hubbard model ladder
with
a controlled RG method
found a phase with a
finite spin-gap and a single gapless charge mode that they interpret as
the analog of a 1D superconductor or a charge density wave.
H. Schulz (Orsay preprint) studied two coupled Luttinger liquids reporting a
spin-gap
in the spectrum and either dominant charge density wave or singlet
pairing correlations in the ground state.

\end